\documentstyle{preprint}

\def\d{\,{\rm d}}

\def\bcn{\begin{center}}
\def\ecn{\end{center}}
\newcommand{\beqn}{\begin{eqnarray}}
\newcommand{\eeqn}{\end{eqnarray}}
\newcommand{\epm}{\mbox{$e^+e^-$}}
\newcommand{\eem}{\mbox{$e^-e^-$}}
\newcommand{\ep}{\mbox{$e^-\gamma$}}
\newcommand{\pp}{\mbox{$\gamma\gamma$}}
\newcommand{\LSP}{\mbox{$\tilde\chi^0_1$}}
\newcommand{\lc}{linear collider}
\newcommand{\bkg}{background}
\newcommand{\sm}{standard model}
\newcommand{\susy}{supersymmetry}
\newcommand{\susic}{supersymmetric}
\newcommand{\mssm}{minimal supersymmetric standard model}
\newcommand{\nwa}{narrow width approximation}
\newcommand{\br}{branching ratio}

\newcommand{\cm}{center of mass}
\newcommand{\xs}{cross section}

\newcommand{\lsp}{lightest supersymmetric particle}
\def\be{\begin{equation}}
\def\ee{\end{equation}}
\def\barr{\begin{array}}
\def\earr{\end{array}}
\def\lsim{\:\raisebox{-0.5ex}{$\stackrel{\textstyle<}{\sim}$}\:}

\def\sel{\mbox{$\tilde e^-$}}
\def\msel{\mbox{$m_{\tilde e}$}}

\def\and{\qquad {\rm and } \qquad}

\def\ie{ {\it i.e.} }
\def\viz{ {\it viz.} }
\def\eg{ {\it e.g.} }
\def\bib{\bibitem}

\def\ib#1,#2,#3{       {\it ibid.\/ }{\bf #1} (19#2) #3}
\def\ap#1,#2,#3{       {\it Ann.~Phys.~(NY)\/ }{\bf #1} (19#2) #3}
\def\ijmp#1,#2,#3{     {\it Int.~J.~Mod.~Phys.\/ } {\bf A#1} (19#2) #3}
\def\mpl#1,#2,#3 {     {\it Mod.~Phys.~Lett.\/ } {\bf A#1} (19#2) #3}
\def\np#1,#2,#3{       {\it Nucl.~Phys.\/ }{\bf B#1} (19#2) #3}
\def\npps#1,#2,#3{     {\it Nucl.~Phys.~B (Proc.~Suppl.)\/ }{\bf B#1}
                             (19#2) #3}
\def\plb#1,#2,#3{      {\it Phys.~Lett.\/ }{\bf B#1} (19#2) #3}
\def\pr#1,#2,#3{       {\it Phys.~Rev.\/ }{\bf D#1} (19#2) #3}
\def\prep#1,#2,#3{     {\it Phys.~Rep.\/ }{\bf #1} (19#2) #3}
\def\prl#1,#2,#3{      {\it Phys.~Rev.~Lett.\/ }{\bf #1} (19#2) #3}
\def\pro#1,#2,#3{      {\it Prog.~Theor.~Phys.\/ }{\bf #1} (19#2) #3}
\def\rmp#1,#2,#3{      {\it Rev.~Mod.~Phys.\/ }{\bf #1} (19#2) #3}
\def\sp#1,#2,#3{       {\it Sov.~Phys.-Usp.\/ }{\bf #1} (19#2) #3}
\def\zpc#1,#2,#3{      {\it Zeit.~f\"ur Physik\/ }{\bf C#1} (19#2) #3}

\begin{document}

\begin{flushright}
{\bf hep-ph/9412245} \\[4ex]
MPI-PhT/94-86\\
December 1994
\end{flushright}

\vskip2cm

\begin{frontmatter}
\title{Production of Heavy Selectrons in \ep\ Collisions}
\author{Debajyoti Choudhury}
\address{{\tt debchou@surya11.cern.ch}\\
        Theory Division,
        CERN,
        CH--1211 Gen\`eve 23,
        Switzerland}
\author{Frank Cuypers}
\address{{\tt cuypers@iws166.mppmu.mpg.de}\\
        Max-Planck-Institut f\"ur Physik,
        Werner-Heisenberg-Institut,
        F\"ohringer Ring 6,
        D--80805 M\"unchen,
        Germany}
\begin{abstract}
We study the production and decay of heavy selectrons
in the \ep\ mode of a \lc\ of the next generation.
The \sm\ \bkg s
can be substantially reduced by appropriate kinematical cuts.
As a consequence,
selectrons far heavier than the kinematical threshold for pair production
are shown to be easily discoverable
for large portions of the \susy\ parameter space.
We also describe a model-independent
kinematical measurement of the mass of the lightest neutralino.
\end{abstract}
\end{frontmatter}
\clearpage

\section{Introduction}

One of the most promising ideas that take us beyond the
\sm\ is \susy\ \cite{susy}.
However, while it cures many of the ills that plague
the \sm, it has its own attendant problems, not the least of
which is the fact that, till
date, we have seen no evidence for its existence. This naturally sets a
minimum scale for the breaking of this symmetry. A second problem deals with
the proliferation of parameters in such a theory, especially in the context
of the soft supersymmetry breaking terms that must be introduced. Though
several attempts \cite{low-energy}
have been made to constrain the parameter space by demanding consistency with
the experimentally measured rates for low--energy processes, such bounds are
most often easily evaded if the mass splittings between supersymmetric
particles of a given kind are comparatively small. In addition, such
constraints have the drawback of being of an indirect nature and hence
dependent on cancellations (or lack thereof) between the contributions from
various diagrams with different virtual exchanges.

A complementary (or even better) method would be to look for an actual
production of the superpartners in high energy collisions.
Of course,
the direct production of new particles is only possible
provided the latter are sufficiently light.
Typically,
a collider facility requires a \cm\ energy
comfortably in excess of twice the mass of the particles to be discovered
to permit their pair-production.
Consequently,
the tightest bounds on the weakly interacting \susic\ sector
are at present given by the LEP experiments
and come close to $m_Z/2$.

In this article, we aim to perform
an analysis in the context of the linear
colliders which are currently being planned,
such as CLIC, JLC, NLC, TESLA, VLEPP, {\it etc.}.
To be specific,
we concentrate on the ``canonical'' design
with a \cm\ energy $\sqrt{s}=500$ GeV,
an electron beam polarization of 90\%
and an integrated luminosity of 10 fb$^{-1}$.
Though most often these machines are thought of
in terms of \epm\ facilities,
they can also be transformed to function in
the \eem, \pp\ and \ep\ modes.
The last-mentioned is the one which is most suitable for our
purpose as it lends itself to
single selectron production,
circumventing in this way the kinematic bound that pair production entails
in \epm, \eem\ and \pp\ collisions.

The production of a single selectron
in association with a neutralino
was analyzed in Refs~\cite{eg,bbc}.
We improve these studies here
by taking fully into account the gaugino mixings,
considering polarized electron and photon beams
and performing a detailed analysis
of the differential \xs s of the signal and its \sm\ \bkg s.
The latter allows us to devise very efficient kinematical cuts
which dramatically enhance the signal to \bkg\ ratio.
As a result,
the \susy\ parameter space
can be explored much deeper
than previously expected \cite{eg}.

In the next Section
we describe the production and decay of a selectron
in \ep\ scattering,
as well as the dominant \sm\ \bkg s.
In Section 3
we explain how photon beams can be obtained at a \lc.
We study in the following Section
the kinematical characteristics of the event distributions
for the signal and \bkg\ reactions.
Armed with this knowlegde,
we analyze in Section 5
which regions of the \susy\ parameter space
can be explored with the machine described above.
Finally,
we summarize our results in the Conclusion.

\section{Signal and Backgrounds}

We restrict ourselves to the \mssm\  with unbroken $R$--parity.
In most realistic scenarios, the \lsp\ is the lightest neutralino \LSP,
which is stable and hence escapes detection.
We concentrate here on the reaction
\be
e^- \gamma \longrightarrow \sel \LSP
\ ,   \label{prodn}
\ee
and the subsequent selectron decay
\be
\sel \longrightarrow e^- \LSP
\ .     \label{decay}
\ee
The \br\ for this decay can be as high as nearly 100\%,
but it can also be substantially lower
if other two body decay modes
are kinematically allowed:
\beqn
 \sel & \longrightarrow & e^- \tilde\chi_i^0 \qquad (i=2,3,4) \ , \\
 \sel & \longrightarrow & \nu_e \tilde\chi_i^- \qquad (i=1,2) \ ,
\eeqn
where the $\tilde\chi_i^0$ and $\tilde\chi_i^-$
are any of the neutralinos or charginos in the theory.

Obviously both the cross section (\ref{prodn}) and the partial
decay rate (\ref{decay})
depend on the details of the supersymmetric extension,
particularly the masses and the mixing angles in the chargino and neutralino
sectors. These receive contributions not only from terms originating in
the naive supersymmetrization of the \sm\ lagrangian, but also from soft
supersymmetry breaking terms. The chargino  mass matrix, in the
($\tilde{W}^- \ \tilde{h}^- $)  basis is  given by \cite{susy}
\be
M_- = \pmatrix{M_2  & \sqrt{2} M_W c_\beta \cr
                \sqrt{2} M_W s_\beta & \mu \cr}
\ ,  \label{chargino}
\ee
whereas the neutralino mass matrix, in the  ($\tilde{\gamma}^0 \
\tilde{Z}^0 \ \tilde{h}_1^0 \ \tilde{h}_2^0 $) basis, is
\be
M_0 = \pmatrix{ M_1 c_w^2 + M_2 s_w^2 & (M_1 - M_2) c_w s_w  & 0 & 0 \cr
                 (M_1 - M_2) c_w s_w  & M_1 s_w^2 + M_2 c_w^2
                                       &   m_Z s_\beta &  - m_Z c_\beta \cr
                0 &  m_Z s_\beta & 0 & - \mu \cr
                0 &  - m_Z c_\beta & - \mu & 0\cr}
\ .      \label{neutralino}
\ee
Here $M_{1,2}$ are the soft $U(1)_Y$ and $SU(2)_L$ \susy\ breaking masses
which parametrize direct gaugino mass terms,
whereas $\mu$ is the higgsino mixing mass.
Furthermore,
$s_w = \sqrt{1 - c_w^2} = \sin \theta_w$
and
$s_\beta = \sqrt{1 - c_\beta^2} = \sin \theta_\beta$
where
$\theta_w$
is the weak mixing angle and
$\tan\beta = v_1/v_2 $
is the ratio of the vacuum expectation values of the two Higgs fields.
We assume the GUT relation
\be
M_1 = \frac{5}{3} M_2 \tan^2 \theta_w
\ ,
\ee
thus leaving us with three parameters namely $\mu,M_2,\tan\beta$.
The exact eigensystem has been computed in Ref.~\cite{guchait}.
We confirm these results
and use them in our calculations.

One further question remains unsettled, though.
It concerns possible mixings between the various charged lepton states.
However, both constraints from low energy  experiments \cite{soft} and
theoretical studies \cite{jan} suggest that such mixings are very small. We
shall thus assume these to be absent.
Also, for simplicity, we regard
$\tilde{e}_L^-$ and $\tilde{e}_R^-$ to be degenerate,
with mass \msel.

The signal we focus on
is thus a final state comprising just a single electron
associated with missing energy and momentum:
\be
e^- \gamma \longrightarrow \tilde e^- \LSP \longrightarrow e^- \LSP \LSP
\ .    \label{signal}
\ee
A considerable background exists,
though, on account of the \sm\ resonant processes
\be
e^- \gamma  \longrightarrow  e^- Z^0  \longrightarrow
         e^- \nu_i \bar{\nu}_i
\label{Z-bg}
\ee
and
\be
e^- \gamma  \longrightarrow  W^- \nu_e  \longrightarrow
         e^- \nu_e \bar{\nu_e}
\ .   \label{W-bg}
\ee
We neglect non-resonant processes
which also contribute to the same observable signature
\be
e^- + \gamma \longrightarrow e^- + \nu_i + \bar{\nu}_i \ .
   \label{bkgd}
\ee
where $i$ is a generation index.
This is an excellent approximation \cite{aco}
which provides a clear picture
of the kinematic distribution of the final state.
We can safely use the \nwa\
to compute the total and differential \xs s
of the processes (\ref{signal},\ref{Z-bg}),
since the decaying particle is either a scalar
or it decays invisibly.
The subprocess \xs s for these reactions
have been first calculated by Renard \cite{renard}.
They were also calculated for particular helicity combinations
in Ref.~\cite{mourao}.
The expressions in Ref.~\cite{bbc} carry some minor errors.

For the process (\ref{W-bg}),
though,
the spin-angle correlations have to be taken carefully into account,
because the final state electron
originates from the decay of a {\em vector} boson.
To do this,
we have treated this reaction as a resonant $2\to3$ process.
The analytic expressions for the differential \xs s
are too long to be displayed here.

Before embarking on further calculations,
it is worth noting
that the \bkg\ process (\ref{W-bg})
can of course be very simply reduced with polarized electron beams.
Unless stated otherwise,
we shall consider in the following
a 90\%\ right-polarized electron beam.

\section{Photon Beams}

High energy photon beams can be obtained
by back-scattering a laser ray
off a high energy electron beam \cite{ginzburg}.
The result of this Compton scattering is that
the electrons are deflected and dumped
while the photons are boosted
into a hard collimated beam.
If the laser is sufficiently intense,
all electrons have a chance to interact,
so that there is no loss in luminosity.

\begin{figure}[htb]
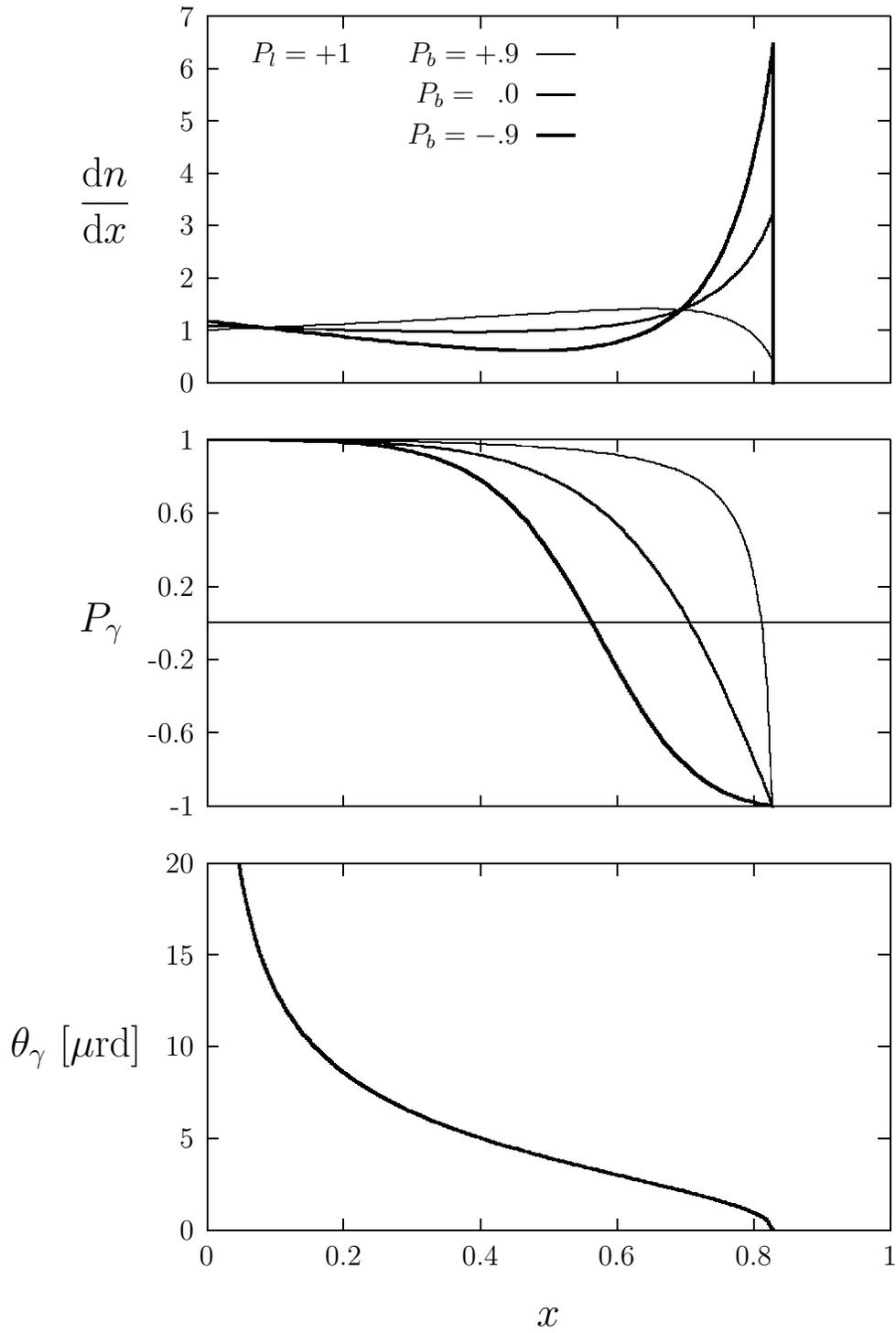

\centerline{\input eny.tex}
\vspace{-1.5cm}
\centerline{\input pol.tex}
\vspace{-1.5cm}
\centerline{\input ang.tex}
\caption{Back-scattered photon energy spectrum,
        polarization and polar angle
        as functions of $x=E_\gamma/E_b$
        for three different combinations of beam polarizations.}
       \label{photon}
\end{figure}

Consider laser photons of energy $E_l$
(${\cal O}(1)$ eV)
and circular polarization $P_l$
scattering with electrons of energy $E_b$
(${\cal O}(10^{11})$ eV)
and longitudinal polarization $P_b$.
(In the next sections we will also need to define the polarization $P_e$
of the the other electron beam,
which is not Compton-converted.)
If the angle between the laser and electron beams be $\theta_{bl}$,
the scattering may be parametrized in terms of
\be
 z  =  \displaystyle
      \frac{4 E_b E_l }{m_e^2} \cos^2 \frac{\theta_{bl}}{2} \ ,
\ee
where $m_e$ is the mass of the electron.
For large $z$, multiple
scattering (\eg, pair creation in conjunction with a laser photon) becomes
important \cite{telnov}
and this results in a depletion of the high energy end of the spectrum.
To prevent this, one needs to ensure that
\be
 z       \leq 2 ( 1 + \sqrt{2} )
\ . \label{ymax}
\ee
In the following we shall assume the equality in the above expression.
The resultant photon normalized energy spectrum $n(x)$
and polarization $P_\gamma(x)$
are depicted in Figs~\ref{photon}
as functions of the fraction $x = E_\gamma/E_b$
of the electron energy carried by the photon.
The underlying analytic expressions are \cite{ginzburg}
\beqn
{\d n(x) \over \d x} =
\frac{1}{\cal N}
& \Biggl\{ & 1 - x + \frac{1}{1 - x} - \frac{4 x }{z ( 1 - x) }
                           + \frac{4 x^2 } {z^2 (1 - x)^2} \nonumber \\
& + & P_b P_l \frac{x (2 - x)}{1 - x}
                       \left[ \frac{2 x} {z (1 - x)} - 1 \right]
\Biggr\}
\ , \label{spec}
\eeqn
\be
P_\gamma (x) = \frac{P_l \zeta (2 - 2 x + x^2)  + P_b x ( 1 + \zeta^2)}
               { (1 -x ) (2 - 2 x + x^2) - 4 x ( z - z x - x)/z^2
                         - P_b P_l \zeta x ( 2 - x) }
\ , \label{polar}
\ee
where
\be
0 \leq x \leq \frac{z}{z + 1}
\ , \label{e101}
\ee
\beqn
{\cal N}
& = &
        \frac{z^3 + 18 z^2 + 24 z + 8}{2 z (z + 1)^2 }
        + \left( 1 - \frac{4}{z} - \frac{8}{z^2} \right) \ln (1 + z)
\nonumber \\
& + &
        P_b P_l \left[ 2 - \frac{z^2}{(z + 1)^2}
        - \left( 1 + \frac{2}{z} \right) \ln (1 + z) \right]
\ , \label{e102}
\eeqn
\be
\zeta = 1 - x ( 1 + 1/z)
\ . \label{e103}
\ee

By design,
the energy of the photons can never reach the beam energy
since their spectrum is limited by Eq.~(\ref{ymax},\ref{e101}).
Furthermore,
in practice the low energy tail of the photon spectrum
cannot participate in any reaction either.
Indeed,
as is displayed in the last of Figs~\ref{photon},
there is a one to one relationship
between the energy of the back-scattered photons
and their angle with respect to the direction of the initial electron:
harder photons are emitted at smaller angles
whereas softer photons are emitted at larger angles.
For small deflection angles from the beam direction we have
\be
\theta_\gamma (x) \simeq {m_e \over E_b} \sqrt{{z \over x}-z-1}
\ee
Clearly,
the photons will be distributed according to an effective spectrum,
which effectively throws  out the low energy photons,
these being produced at too wide an angle to contribute significantly
to any reaction.
The exact profile of this effective spectrum
depends somewhat on the shape of the electron beam.
In the absence of a detailed
(and machine specific)
study of this effect,
we approximate this effect
by a sharp cut.
The position of this cut
depends of course crucially on the conversion distance,
{\em i.e.} the distance between the interaction point
and the point where the laser photons are back-scattered.
For example, if we assume a conversion distance of 5 cm
and an interaction spot of 500 nm diameter,
then photons scattered at more than 5 $\mu$rd are lost.
The lowest energy photons would then have
$x_{\rm min} \approx .4$,
as can be read from the last of Figs~\ref{photon}.
In the following we use ({\em cf.} Eq.~(\ref{ymax}))
\be
x_{\rm min} = .5
\qquad
x_{\rm max} = \frac{ 2+2\sqrt{2}}{ 3+2\sqrt{2}} \approx 0.8284
\ . \label{ymin}
\ee
We have checked that our final results
are not very sensitive
to the choice of the first of these two machine parameters.
Any \xs\ is obtained
by convoluting the fixed energy scattering
\xs\ with the photon spectrum (\ref{spec}):
\be
\d\sigma(s) = \int_{x_{\rm min}}^{x_{\rm max}} \d x \
{\d n \over \d x} \d\sigma(x s)\
\theta\left(x - \frac{1}{s} \left(\sum_i m_i\right)^2\right)
\ ,
\ee
where the sum runs over the masses of the final state particles.

\section{Kinematical Cuts}

In the absence of linear polarization,
the final state electrons
in either the signal or the background processes
are characterized by only two independent kinematical variables,
say the cosine of their polar angle ($\cos\theta_e$)
and their energy ($E_e$).
As it turns out,
the distribution of events in this plane
is very different for the three processes (\ref{signal}--\ref{W-bg}).
We now proceed to study the boundaries of these distributions
in order of increasing complexity.

\subsection{$e^-\gamma \to e^-Z^0$}
Were the initial state photons monochromatic
($E_\gamma=\sqrt{s}/2$),
the final state electrons would be distributed
in the $(\cos\theta_e,E_e)$ plane
along the line of constant energy
$E_e=(s-m_Z^2)/2\sqrt{s}$.
Since the energy of the initial state photon
is actually spread between
$x_{\rm min}\sqrt{s}/2 \leq E_\gamma \leq x_{\rm max}\sqrt{s}/2$
(\ref{ymin}),
the final state electrons are distributed in phase space region
\begin{equation}
\cos\theta_e
=
{(1+x)\sqrt{s}E_e - xs + m_Z^2 \over (1-x)\sqrt{s}E_e}
\qquad
x_{\rm min}\leq x\leq x_{\rm max}
\ .\label{e1}
\end{equation}
The two boundary curves are displayed in Figs~\ref{scatr} and \ref{scatl}.
They confine all the $Z^0$ events,
and will be used as kinematical cuts
in the subsequent numerical analysis
to entirely eliminate the $Z^0$ \bkg\ (\ref{Z-bg}).

\subsection{$e^-\gamma \to \tilde e^- \tilde\chi_1^0 \to e^- \tilde\chi_1^0
\tilde\chi_1^0$}
The electrons emerging from the decay of the selectron
are distributed within the phase space region
\begin{equation}
E_e
=
{m_{\tilde e}^2-m_{\tilde\chi}^2
\over
2\left[
E_{\tilde e}-k_{\tilde e} \cos\left(\theta_{\tilde e}-\theta_e\right)
\right]}
\ ,\label{e2}
\end{equation}
where $k_{\tilde e}=\sqrt{E_{\tilde e}^2-m_{\tilde e}^2}$ is the momentum of
the selectrons,
whose energy spreads within the region
\begin{equation}
\cos\theta_{\tilde e}
=
{(1+x)\sqrt{s}E_{\tilde e} - xs - m_{\tilde e}^2 + m_{\tilde\chi}^2 \over
(1-x)\sqrt{s}k_{\tilde e}}
\qquad
x_{\rm min}\leq x\leq x_{\rm max}
\ .\label{e3}
\end{equation}
Since the (heavy) selectrons are produced rather isotropically \cite{eg},
the cosine on the left hand side of Eq.~(\ref{e3}) can take any value
between $-1$ and $1$.
As a consequence,
the highest energy of the selectrons is given by
\begin{eqnarray}
E_{\tilde e}^{\rm max}
=
{1 \over 4x_{\rm max}\sqrt{s}}
&\Biggl[&
(1+x_{\rm max}) \left(x_{\rm max}s+m_{\tilde e}^2-m_{\tilde\chi}^2\right)
\nonumber\\
&+&
(1-x_{\rm max})
        \sqrt{ \left(x_{\rm max}s+m_{\tilde e}^2-m_{\tilde\chi}^2\right)^2
                - 4x_{\rm max}sm_{\tilde e}^2 }
\quad\Biggr]
\label{e4}
\end{eqnarray}
and the energy of the decay electrons is confined within
\begin{equation}
{m_{\tilde e}^2-m_{\tilde\chi}^2
\over
2\left(
E_{\tilde e}^{\rm max} + k_{\tilde e}^{\rm max}
\right)}
\leq E_e \leq
{m_{\tilde e}^2-m_{\tilde\chi}^2
\over
2\left(
E_{\tilde e}^{\rm max} - k_{\tilde e}^{\rm max}
\right)}
\ .\label{e5}
\end{equation}
The two boundary lines are displayed in Figs~\ref{scatr} and \ref{scatl}
for a selectron of mass $\msel=250$ GeV
and the particular set of \susy\ parameters
$\mu=500$ GeV,
$M_2=400$ GeV and
$\tan\beta=4$.
For this choice,
the mass of the lightest neutralino is
$m_{\tilde\chi^0_1}=193$ GeV.
The available phase space is thus small
because the sum of the produced masses
$\msel+m_{\tilde\chi^0_1}=443$ GeV
is so close to the available \cm\ energy
$\sqrt{x_{\rm max}s}=455$ GeV.

Note that the values of the two energy boundaries (\ref{e5})
depend solely on
the \cm\ energy,
the mass of the selectron and
the mass of the neutralino.
Therefore,
a measurement of these two limiting energies
provides a direct kinematical determination
of the mass of the neutralino.
A similar model independent measurement of the neutralino mass
can otherwise be performed only in a polarized $e^-e^-$ experiment,
if the selectron is light enough to be pair-produced \cite{ee}.

\subsection{$e^-\gamma \to W^-\nu_e \to e^-\bar\nu_e\nu_e$}
Here the situation is more subtle.
At first glance
one might expect the decay electrons of the $W^-$
to be subjected to similar kinematical bounds
as the decay electrons of the selectrons.
Indeed,
proceeding with the replacements
$m_{\tilde e} \leftrightarrow m_W$ and
$m_{\tilde\chi} \leftrightarrow m_\nu = 0$
in Eqs~(\ref{e2},\ref{e5}),
the energy of the $W^-$ decay electrons is in principle only bound by
\begin{equation}
{m_W^2 \over 2\sqrt{s}}
\leq E_e \leq
{\sqrt{s} \over 2}
\ .\label{e6}
\end{equation}
This range covers almost the entire phase space.
However,
some improvement can be achieved
by identifying the dense regions in phase space.
Since the $W^-$ production cross section
is very much peaked backwards
with respect to the initial electron beam \cite{eg},
it is a good approximation to set $\cos\theta_W=-1$ in
Eqs~(\ref{e2},\ref{e3}).
Hence,
the most probable energy of the $W^-$ is
\begin{equation}
E_W
=
{x^2s+m_W^2 \over 2x\sqrt{s}}
\qquad
x_{\rm min}\leq x\leq x_{\rm max}
\ ,\label{e7}
\end{equation}
and the bulk of the electrons is produced
within the phase space region
\begin{equation}
E_e
\leq
{m_W^2
\over
2\left( E_W+k_W \cos\theta_e \right)}
\ .\label{e8}
\end{equation}

\begin{figure}[htb]
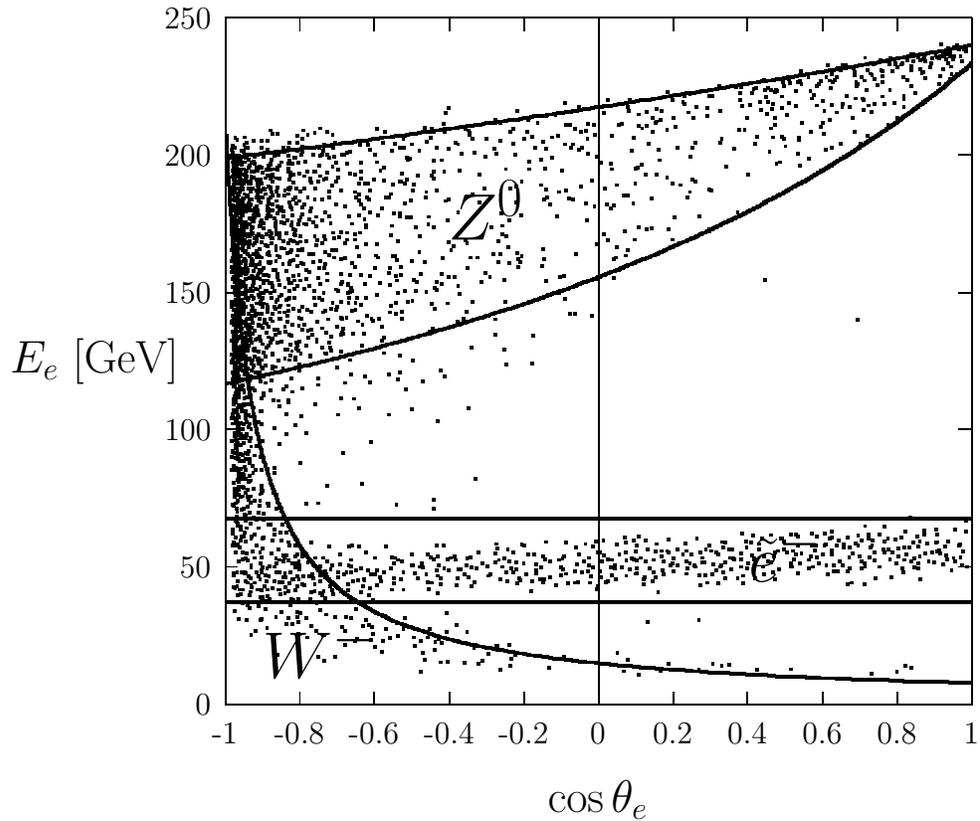

\centerline{\input scat_r.tex}
\caption{Scatter plot showing the angle--energy distributions for the
        final state electrons
        emanating from the processes
        (\protect\ref{signal},\protect\ref{Z-bg},\protect\ref{W-bg}).
        Assuming an integrated luminosity of 10 fb$^{-1}$
        each points corresponds to one event.
        The curves show the exact ($Z^0,\sel$)
        or approximate ($W^-$)
        kinematically allowed ranges.
        The laser photons are 100\%\ left polarized
        whereas both electron beams are 90\%\ right polarized
        ($P_l=-1 \quad P_b=+.9 \quad P_e=+.9$).}
       \label{scatr}
\end{figure}

\begin{figure}[htb]
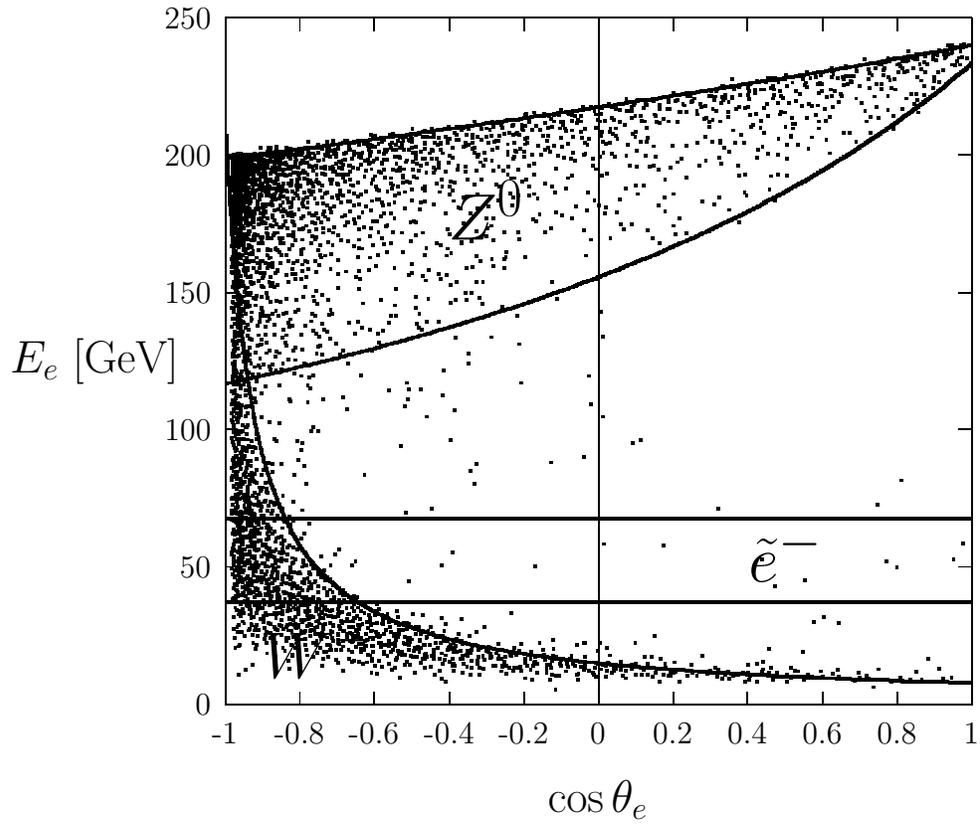

\centerline{\input scat_l.tex}
\caption{Same as Fig.~\protect\ref{scatr},
        except for the laser and Compton-converted electron beams
        which have now opposite polarizations
        ($P_l=+1 \quad P_b=-.9 \quad P_e=+.9$).}
       \label{scatl}
\end{figure}

Since the $W^-$ production \xs\
increases monotonically with the \cm\ energy \cite{eg,ginz2},
more events occur for larger values of $x$.
In Figs~\ref{scatr} and \ref{scatl},
we have thus displayed the curve corresponding to $x=x_{\rm max}$
in Eqs~(\ref{e7},\ref{e8}).
Clearly,
a large portion of the $W^-$ events is confined below this curve.
It will therefore be used as a kinematical cut
alongwith Eq.~(\ref{e1})
in the subsequent numerical analysis.

It should be borne in mind that
events which do not satisfy this bound
are also kinematically allowed.
However,
the corresponding rate is strongly suppressd by the dynamics
of the $W^-$ production mechanism.
One effect of the dynamics can be seen
by comparing the density of events plotted on
Figs~\ref{scatr} and \ref{scatl}.
The only difference consists of different initial state polarizations.
Although the choice of Fig~\ref{scatl}
confines the $W^-$ events better within the bounds (\ref{e7},\ref{e8}),
far fewer selectrons are produced
and the signal to \bkg\ ratio is worse.
This is why we concentrate in the next section
on the choice of polarizations used in Fig~\ref{scatr}.

\section{Results}

In addition to the kinematical cuts (\ref{e1},\ref{e8}) discussed above
and depicted in Figs~\ref{scatr} and \ref{scatl},
the signal to \bkg\ ratio can be further enhanced
by dividing the $(\cos\theta_e,E_e)$ phase space into $N$ bins
(of equal size, here).
The number of events in each bin
is then compared with the \sm\ expectation.
This procedure takes automatically into account
the information contained in Eq.~(\ref{e5}).
The significance of the deviation is given by the $\chi^2$ test
\be
\chi^2 = \sum_i^N
\left(
        {n^{\rm exp}-n^{\rm SM} \over \Delta n^{\rm exp}}
\right)^2
\ , \label{chi}
\ee
where $n^{\rm SM}$ is the number of events expected for the \sm,
$n^{\rm exp} = n(\mu,M_2,\tan\beta,\msel)$
is the corresponding observed number of events
(if \susy\ is to explain the deviation)
and the error $\Delta n$
is the quadratic combination of statistical and systematic
errors on the observed number of events
\be
\Delta n = \sqrt{n + (\epsilon n)^2}
\ .\label{err}
\ee
The relative systematic error $\epsilon$
is essentially due to the luminosity measurement
(the uncertainty on the electrons' energies and angles is negligible)
and is set to 1\%\ in the following.

Clearly,
if less than five events are contained in a bin,
the probabilistic interpretation of the $\chi^2$ test
becomes unreliable.
Indeed,
with so few events the underlying Poisson distribution
does not resemble enough a gaussian shape
to warrant the sum (\ref{chi}) to be distributed according to a $\chi^2$.
For this reason,
we ignore alltogether any bin with less than five events.

Instead of attempting to explore at one go
the four-dimensional parameter space \viz
($\mu,M_2,\tan\beta,\msel$),
we choose to present our results in the form
of 2-dimensional $\chi^2$ contour plots
in the $(\mu,M_2)$ plane
for different values of $\msel$ and $\tan\beta$.
We also exhibit the dependence on
the polarization of the electron beams $P_b$ and $P_e$,
and on the number of bins $N$.
To obtain the explorable regions
at the 95\%\ confidence level
we set $\chi^2=6$ in Eq.~(\ref{chi}).

Unless stated otherwise we use in the following
\be
\msel=250\ {\rm GeV} \qquad
\tan\beta=4 \ , \label{beg}
\ee
\be
P_l=-1\equiv100\%\ {\rm left} \quad
P_b=.9\equiv90\%\ {\rm right} \quad
P_e=.9\equiv90\%\ {\rm right} \ ,
\ee
\be
N=3\times3 \ . \label{end}
\ee

\begin{figure}[htb]
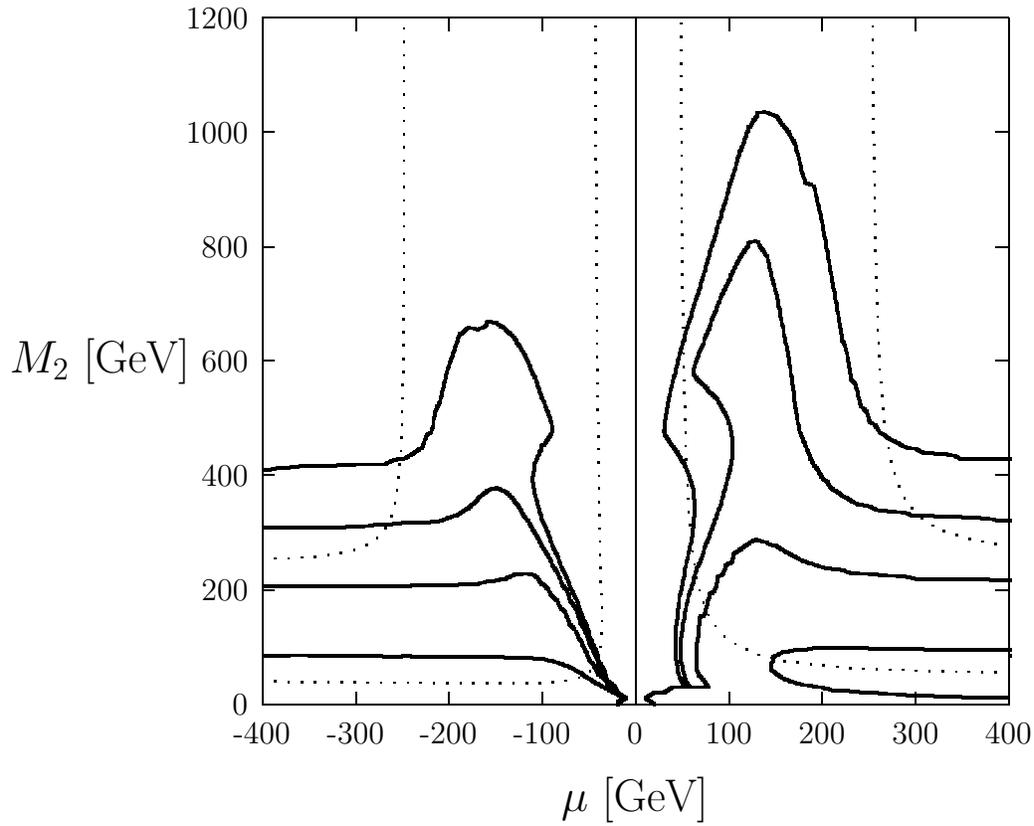

\centerline{\input sel.tex}
\caption{Contours of $\chi^2=6$
        for selectron masses $\msel=250,300,350,400$ GeV
        (from upper to lower curves).
        All other parameters are specified in
        Eqs~(\protect\ref{beg}--\protect\ref{end}).
        The areas below the plain curves can be explored with 95\%\ confidence.
        The dotted curves delimit the region already excluded by LEP I
        (lower curves)
        and the region below which charginos can be pair-produced
        at the same \lc\ run in the \epm\ mode
        (upper curves).}
       \label{sel}
\end{figure}

Since we consider a machine operating with a \epm\ \cm\ energy
$\sqrt{s}_{\rm max}=500 $ GeV,
selectrons with a mass up to 250 GeV
can in principle be pair-produced and detected
in the \epm\ or \eem\ colliding modes.
We therefore display in Fig.~\ref{sel}
observability contours for $\msel \geq 250$ GeV,
\ie {\sl beyond the kinematical limit of the parent collider}.
As expected,
the explorable area shrinks with increasing selectron mass.
The sharp drop for small $\mu$ can easily be understood as for such values,
the \lsp\ is primarily a higgsino
and (almost) does not couple to the electron.
This part of the parameter space can be better explored
using the \epm\ mode to observe chargino pair-production.
The regions of parameter space located between the dotted lines
are those which could be explored this way
at the same collider
(or have already been clearly excluded by the LEP experiments)
assuming charginos can be observed
up to the very threshold for their pair-production.
If this limit can indeed be reached in practice,
the \ep\ mode has a chance of {\em discovering} \susy\
where the \epm\ mode cannot,
only if the selectron mass is comprised within
$250<\msel<325$ GeV,
and this in a marginal portion of the \susy\ parameter space.
Still,
even if \susy\ has been discovered before,
for a 500 GeV machine,
the \ep\ collider operating mode is the only one susceptible
of discovering a selectron heavier than 250 GeV.
Even a selectron as heavy as 400 GeV
can be observed and studied.

\begin{figure}[htb]
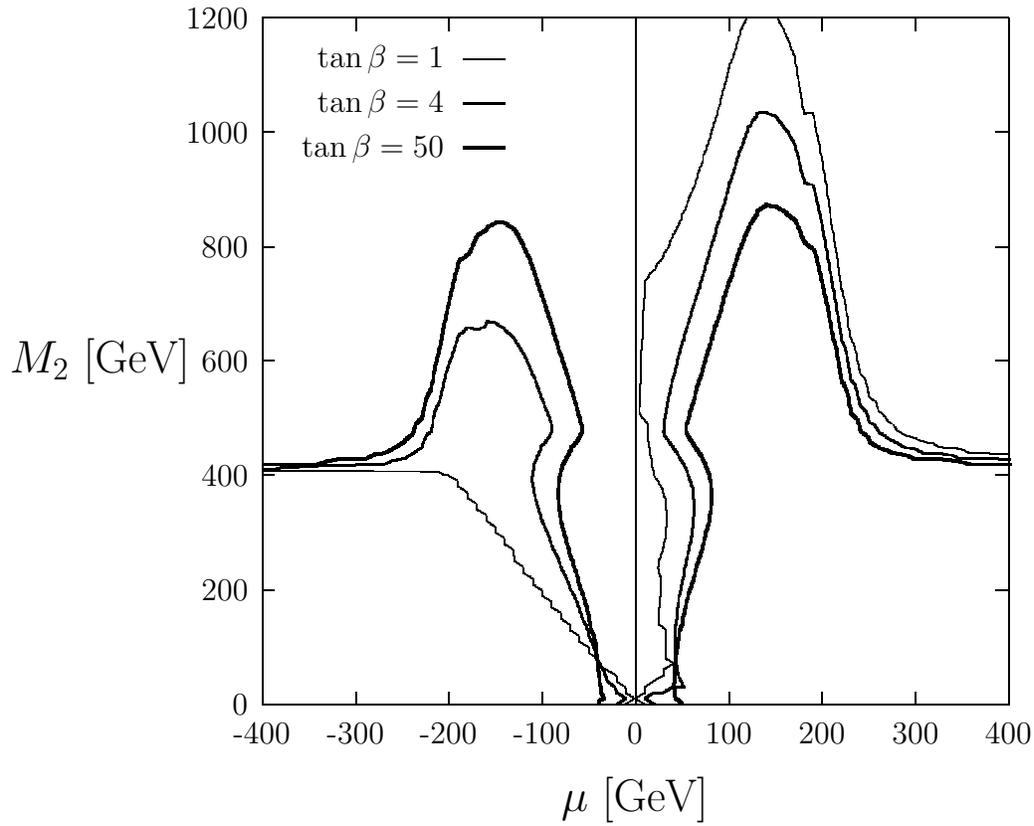

\centerline{\input tan.tex}
\caption{Contours of $\chi^2=6$
        for $\tan\beta=1,4,50$.
        All other parameters are specified in
        Eqs~(\protect\ref{beg}--\protect\ref{end}).
        The areas below the curves can be explored with 95\%\ confidence.}
       \label{tan}
\end{figure}

In Fig.~\ref{tan} we display the $\tan\beta$ dependence.
As is easily evinced,
the contours tend to become more symmetric for progressively
larger values of $\tan\beta$.
The reason can be understood by examining the characteristic equation
of the matrix (\ref{neutralino}).
In the limit where $\tan\beta=\infty$
there remains only a $\mu^2$ dependence.
Similarly,
for $|\mu| \gg m_Z $ the dependence on $\tan\beta$ should vanish.
This fact is indeed reflected by the convergence of the contours
for large values of $|\mu|$.

\begin{figure}[htb]
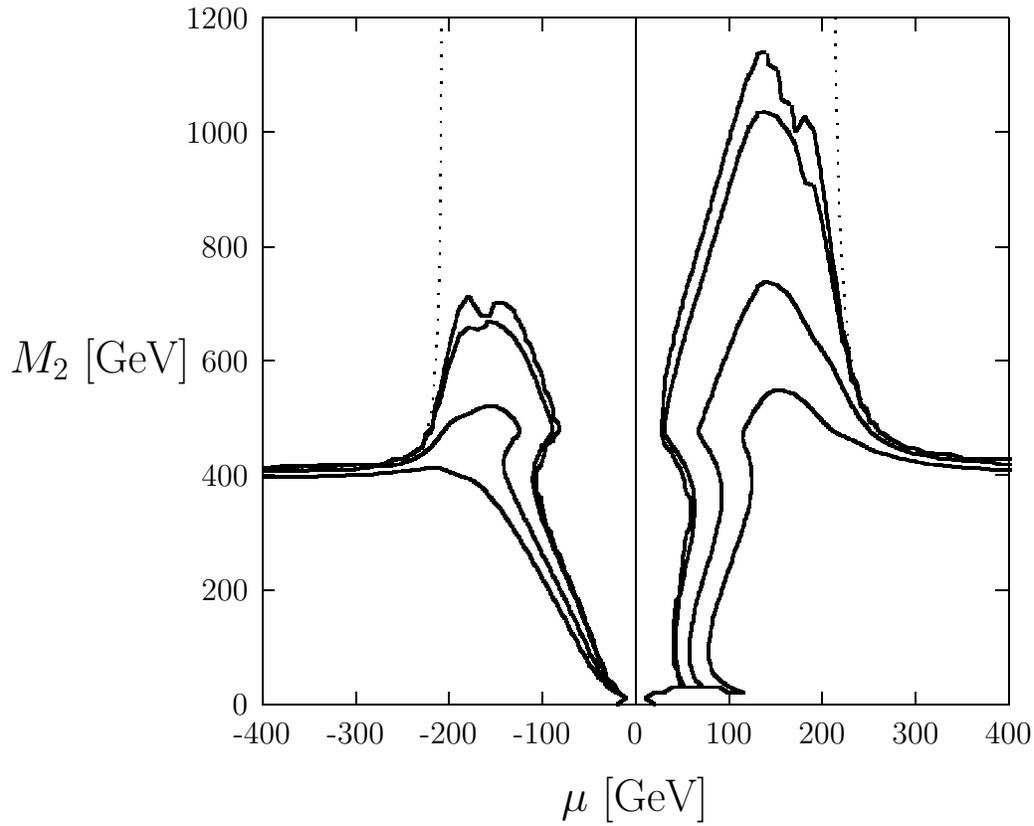

\centerline{\input poldep.tex}
\caption{Contours of $\chi^2=6$
        for electron beam polarizations $P_b=P_e=0,50,90,95\%$
        (from lower to upper curves).
        All other parameters are specified in
        Eqs~(\protect\ref{beg}--\protect\ref{end}).
        The areas below the curves can be explored with 95\%\ confidence.
        The dotted curves depict the kinematical limit
        of the \susic\ reaction (\protect\ref{signal}).}
       \label{poldep}
\end{figure}

Some portion of the $W$ \bkg\ (\ref{W-bg}) being irreducible,
it is important to reduce it as much as possible
by polarizing the initial electron beam.
As can be inferred from Fig.~\ref{poldep}
the incidence of a poor polarization
is particularly dramatic in the region
$ 50 \:{\rm GeV} \lsim \mu \lsim 200 \:{\rm GeV} $,
where the \susic\ signal (\ref{signal}) is small
on account of the \lsp\ being primarily a higgsino.
Nevertheless,
improving the polarization beyond 90\%\
only yields marginal improvements.
This is because
even though the \bkg\ is further decreased,
the signal remains at the limit of observability.
The erratic features of the curve at 95\%\ polarization
reflect our rejection of bins containing less than five events.

\begin{figure}[htb]
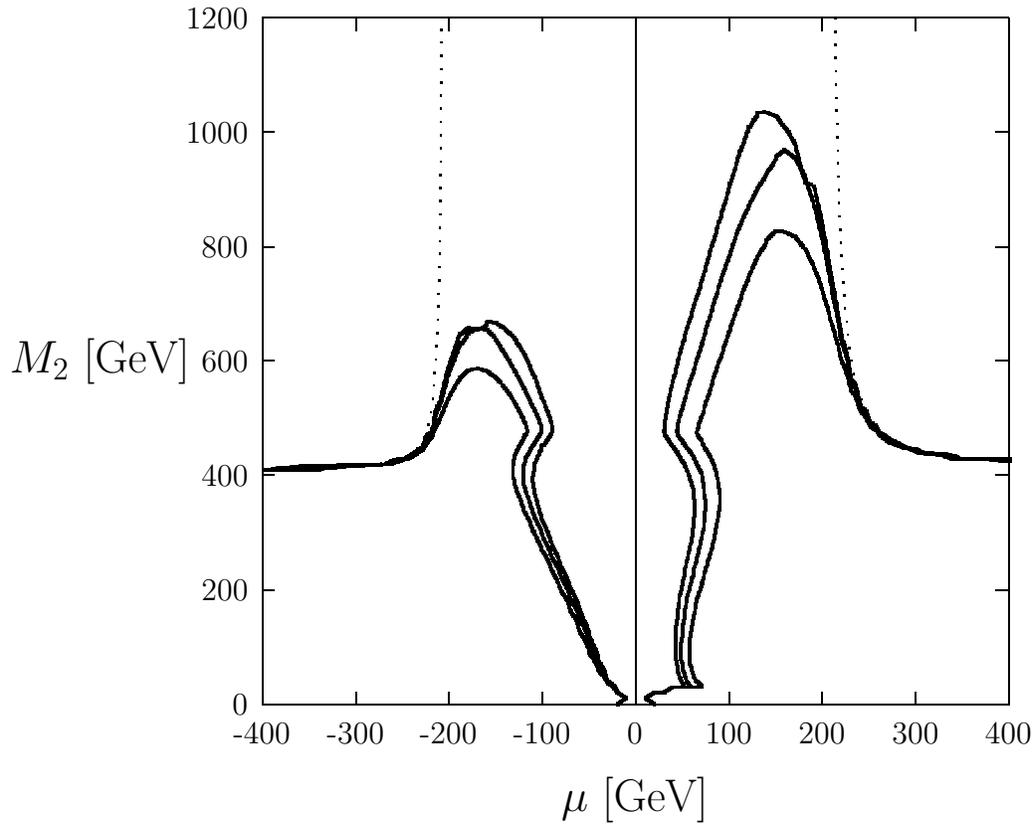

\centerline{\input bin.tex}
\caption{Contours of $\chi^2=6$
        for a number of bins $N=1\times1,2\times2,3\times3$
        (from lower to upper curves).
        All other parameters are specified in
        Eqs~(\protect\ref{beg}--\protect\ref{end}).
        The areas below the curves can be explored with 95\%\ confidence.
        The dotted curves depict the kinematical limit
        of the \susic\ reaction (\protect\ref{signal}).}
       \label{bin}
\end{figure}

Finally,
in Fig.~\ref{bin} we turn to the dependence of our results on the binning,
more specifically on the number of bins.
Since coarser binning loses information about the differential distribution,
the improvement in the bounds
with better energy and angular resolution
is not unexpected\footnote{
This improvement is not a monotonic function of bin cardinality,
as very fine binning would leave too few events in each,
thus disqualifying them from contributing to the $\chi^2$ function.
As a matter of fact,
a $4\times4$ binning already degrades the resolution!}.
It should be noted, though,
that increasing the number of bins
results only in a modest increase in sensitivity.
This is a direct consequence
of the efficient kinematic cuts Eqs~(\ref{e1},\ref{e8}).

The variation of this improvement with $\mu$
can again be traced to the neutralino mass matrix (\ref{neutralino}).
For small $|\mu|$,
the lightest neutralino has a large higgsino component.
As a result,
its coupling to the electron and hence
both the production cross section (eqn.~\ref{prodn}) and the partial
decay width (eqn.~\ref{decay}) are suppressed.
This leads to a comparatively smaller signal to noise ratio.
To offset this loss,
additional information as obtained from binning is useful.
For large values of $|\mu|$ though,
the higgsinos tend to decouple
and the lightest neutralino is primarily a gaugino.
The signal to \bkg\ ratio on imposition of
the kinematic cuts (\ref{e1},\ref{e8}) is sufficiently large
to render binning almost inconsequential.

\section{Conclusions}

We have studied the production and decay of heavy selectrons
at a \lc\ of the next generation operated in its \ep\ mode
in the context of the \mssm.
While the other modes (\epm, \eem\ and \pp)
are limited by the kinematical limit of $\msel<\sqrt{s}/2=250$ GeV,
the \ep\ option can discover selectrons
which are much heavier,
up to 400 GeV.
The \sm\ background can be controlled
by a judicious choice of beam polarizations and kinematical cuts.

The phase space distribution of the signal final state electrons
stands out significantly from the background.
This allows to infer
in a model-independent way
the mass of the selectron
as well as that of the invisible lightest neutralino,
the \lsp.

\clearpage

\clearpage

\clearpage

\clearpage

\clearpage

\clearpage

\clearpage

\end{document}